# Soliton dynamics in microresonators with XPM induced negative thermo-optic effect

Yanzhen Zheng[1], Changzheng Sun[1]*, Bing Xiong[1], Lai Wang[1], Zhibiao Hao[1], Jian Wang[1], Yanjun Han[1], Hongtao Li[1], Yu Jiadong[1], Yi Luo[1], Jianchang Yan[2], Tongbo Wei[2], and Junxi Wang[2]

[1]Beijing National Research Centre for Information Science and Technology (BNRist), Department of Electronic Engineering, Tsinghua University, Beijing 100084, China

[2]R&D Center for Semiconductor Lighting, Institute of Semiconductors, Chinese Academy of Sciences, Beijing 100083, China

KEYWORDS. Nonlinear optics, integrated optics, microcavity, optical frequency comb, aluminum nitride

**ABSTRACT:** Optical frequency comb generation in microresonators has attracted significant attention over the past decade, as it offers the promising potential for chip-scale optical frequency synthesis, optical clocks and precise optical spectroscopy. However, accessing temporal dissipative Kerr soliton (DKSs) is known to be severely hampered by thermal effects. Furthermore, due to the degeneracy of soliton existence range with respect to soliton number, deterministically accessing single soliton state is another challenge. Here, we demonstrate stable and deterministic single soliton generation in AlN-on-sapphire platform via auxiliary laser pumping scheme without the requirement of fast control of the pump power and detuning. Moreover, we reveal the underlying physics of soliton switching in a dual-pumped microcomb, which is fully described by the Lugiato–Lefever equation. The switching process is attributed to cross-phase modulation (XPM) induced degeneracy lifting of the soliton existence range, corresponding to an effective negative thermo-optic effect.

Temporal dissipative Kerr solitons (DKSs) in microresonators correspond to ultra-short optical pulses, which maintain their pulse-shape while circulating inside the microresonators[1]. Generation of DKSs in microresonators paves the way for chip-scale optical spectroscopy[2,3], ultrafast distance measurements[4], optical microwave signal generation[5], optical frequency synthesizer[6], and photonic radar[7,8]. However, accessing steady soliton states is known to be impeded by thermal effects[9], and soliton states are rarely captured by just sweeping the pump laser frequency, especially for microresonators based on materials with high thermo-optic coefficient (e.g. ~ $3.6 \times 10^{-5}$ K$^{-1}$ for AlN and ~ $2.3 \times 10^{-4}$ K$^{-1}$ for AlGaAs). The schemes commonly employed to overcome thermal instability include power-kicking[10], single-sideband modulation[11], and auxiliary laser pumping[12–14]. For many applications, single soliton state is desirable. However, soliton existence range is degenerate with respect to the soliton number in the microresonator[15,16], which hinders deterministic generation of single soliton. Fortunately, the degeneracy of the soliton existence range can be lifted by thermal effects, which allows soliton switching and deterministic generation of single soliton by backward tuning the pump frequency[15].

Here, we report deterministic single soliton generation in a dual-pumped microresonator by cross phase modulation (XPM) induced degeneracy-lifting of soliton existence range. In contrast to thermally induced degeneracy-lifting in microresonators with positive thermo-optic coefficient, XPM shifts the soliton existence range to the red (longer wavelength) side as the soliton number decreases, corresponding to an effective negative thermo-optic effect, thus making it possible to switch between soliton states and eventually access single soliton by forward tuning the pump laser.

First, we experimentally demonstrate soliton switching and single soliton generation in dual-pumped AlN microresonators. AlN-on-sapphire platform (AlNOI)[11,17–19] manifests both second- ($\chi^{(2)}$) and third-order ($\chi^{(3)}$) nonlinearities, as well as a wide transparency window spanning from 0.2 to 13.6 μm[20], making it particularly attractive for chip-scale nonlinear applications ranging from deep ultra-violet (DUV) to mid-infrared (MIR). So far, broadband Kerr and Raman combs have been demonstrated with AlN microresonators[18], yet the high thermo-optic coefficient of AlN makes stable DKS generation a challenge. Single soliton in AlN microresonator was previously captured via single-sideband modulation[11], which requires rapid and sophisticated control of pump laser frequency scan.

The microresonators used in our experiment are fabricated on a 1.2-μm-thick single-crystalline AlN film epitaxially grown on *c*-plane (0001) sapphire by metal organic chemical vapor deposition (MOCVD). The air-clad AlN microring[21] is dispersion engineered to ensure broadband anomalous dispersion. A relatively small radius of 100 μm is adopted to avoid Raman effect from interfering with Kerr comb generation[22],. The optimized cross section of the AlN microring is $3 \times 1.2$ μm$^2$ and the etching depth is 800 nm, as shown in Figure 1(a). The simulated dispersion profiles for TE$_{00}$ and TE$_{10}$ modes are plotted in Figure 1(c). It is noticed that a mode crossing occurs at around 1410 nm, which will lead to dispersive-wave generation at soliton states[23–25]. The measured intrinsic *Q* factors of the fabricated AlN microring are $2.4 \times 10^6$ and $1.4 \times 10^6$ for TE$_{00}$ and TM$_{00}$ modes, corresponding to a waveguide loss of 0.16 dB/cm and

0.28 dB/cm, respectively. The measured and fitted resonance curves for TE$_{00}$ mode are shown in Figure 1(d).

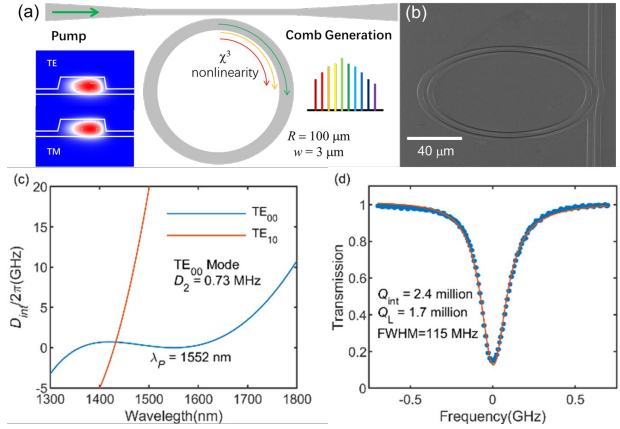

Figure 1 (a) Top view of the AlN microring chip. The inset at the left bottom shows the simulated TE$_{00}$ and TM$_{00}$ modes inside the cavity. (b) Scanning electron microscopy (SEM) image of an AlN microring with a radius of 60 μm. A pulley coupling structure is adopted to enlarge the coupling gap, so as to suppress proximity effect during electron beam lithography (EBL). (c) Simulated dispersion profile assuming a pump wavelength at 1552 nm. Note that a mode crossing occurs at around 1410 nm, which will lead to the generation of dispersive waves. (d) Measured resonance linewidth and extracted Q-factors for TE$_{00}$ mode. The blue dots and the red line correspond to measured and fitted results, respectively.

The experimental setup of the dual-pumping scheme is shown in Figure 2(a). Light from two tunable lasers (TLS1 and TLS2) is coupled into the AlN microring simultaneously. The auxiliary laser (TLS2), which is employed to compensate the thermal effects, is first tuned into resonance and then kept fixed, while the pump laser (TLS1) is swept across the cavity mode from short to long wavelengths (forward tuning) at a relatively low speed (~10 MHz/s). Soliton state can be stably accessed by judiciously adjusting the detuning of the pump and the auxiliary lasers (Supplement 1). In the soliton state, the pump laser is red-detuned with respect to hot cavity mode, whereas the auxiliary laser remains blue-detuned. Frequency comb spectra obtained with different pump and auxiliary laser detuning are shown in Figs. 2(b-d). The comb shown in Figure 2(d) corresponds to the single soliton state, which features a sech$^2$-shape spectral profile. The peak around 1410 nm is attributed to dispersive wave induced by avoided mode-crossing, in agreement with the simulated dispersion profile shown in Figure 1(c).

Once the microresonator is in the DKS state, bi-directional switching of soliton states are observed by forward or backward tuning the pump laser. Figure 2(e) plots the variation of converted comb power as the pump laser is tuned. Forward tuning of the pump results in successive reduction of the soliton number within the microresonator, and single soliton state can be deterministically accessed. This indicates that the soliton existence range degeneracy is lifted in the dual-pumped microresonator. However, the switching dynamics here is different from that reported in [15], where single soliton is accessed via successive reduction of soliton number by backward tuning. In contrast, the soliton number in a dual-pumped microresonator decreases with forward tuning. In addition, spontaneous transition to soliton state is recorded when the pump laser is tuned into resonance from the red-detuned region (Supplement 1). Such self-starting behavior is theoretically predicted for microcombs with negative thermo-optic effect [26], and was previously reported for dual-pumped Si$_4$N$_3$ microresonators[13] or in LiNbO$_3$ resonators with photorefractive effect[27].

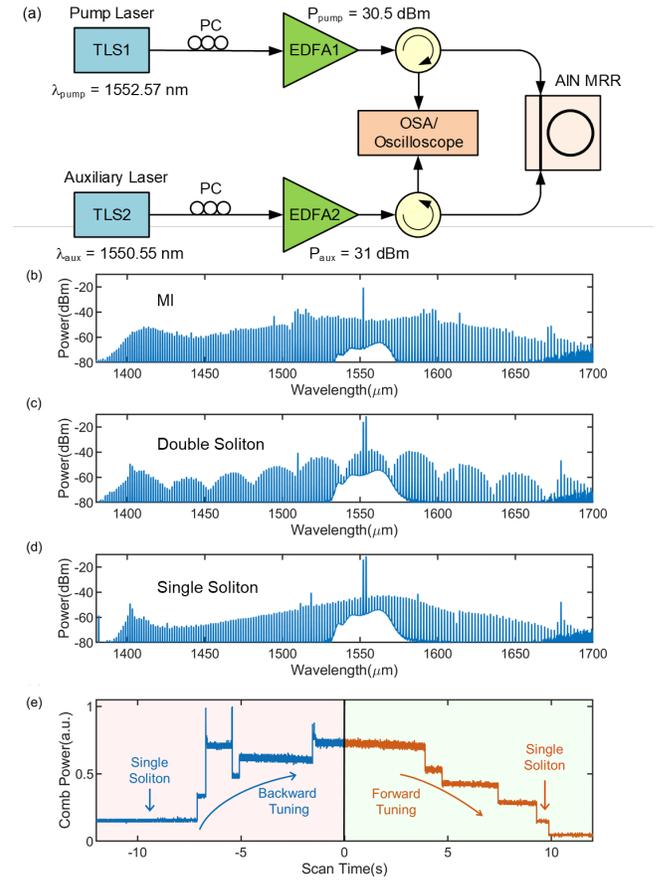

Figure 2 (a) Schematic for the dual-pumping experiment setup. Circulators are employed to separate light entering and exiting the AlN microring resonator (MRR). TLS, tunable laser; EDFA, erbium-doped fiber amplifier; OSA, optical spectrum analyzer; PC, polarization controller. (b-d) Optical spectra of MI comb and soliton combs with both TLS1 and TLS2 set to TE polarization. (b) MI comb, (c) Double solitons, (d) Single soliton. (e) Experimental trace of forward and backward tuning of TLS1. Forward tuning results in successive switching of multi-soliton to single soliton.

To better understand the soliton dynamics in a dual-pumped microresonator, numerical simulations are carried out based on normalized Lugiato-Lefever equations (LLEs) [28,29]

$$\frac{\partial E_i}{\partial t} = -(1+i\delta_i)E_i + i\beta\frac{\partial^2 E_i}{\partial \theta^2} + i\left(|E_i|^2 + 2|E_j|^2\right)E_i + F_i, \quad (1)$$

where subscript $i$ or $j$ denotes either the pump or the auxiliary laser, $E(t,\theta)$ is the normalized optical field traveling in the cavity, $t$ is the normalized slow time associated with the cavity free spectral range (FSR). and $\theta$ is the azimuthal angle along the microring. $\beta$ is the normalized dispersion of the cavity, $F$ is the normalized power of the pump or auxiliary laser, and $i2|E_j|^2E_i$ is the XPM term. $\delta$ is the detuning of the pump or auxiliary laser with respect to the hot cavity mode normalized to the cavity half-linewidth $\Delta\omega$, and can be expressed as $\delta = \tilde{\delta} - \delta_T$, where

$\tilde{\delta} = (\omega_0 - \omega_l)/\Delta\omega$ and $\delta_T = (\omega_0 - \omega_h)/\Delta\omega$ are the absolute detuning and thermal detuning, with $\omega_0$, $\omega_h$ and $\omega_l$ denoting the cold cavity resonance frequency, hot cavity resonance frequency and laser frequency, respectively.

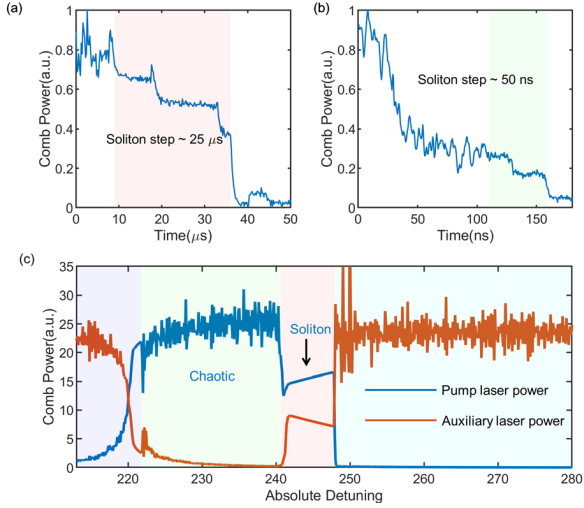

Figure 3 Soliton generation in a dual-pumped AlN microring. Comb power (a) with and (b) without the auxiliary laser at the edge of the thermal triangle recorded under fast pump laser tuning (~20 nm/s). Shaded region in the figure denotes the soliton steps. The soliton step duration is only ~50 ns without the auxiliary laser, but extends to ~25 μs with the introduction of the auxiliary laser. (c) Simulated comb power as a function of pump laser detuning. The total intracavity power is nearly unvaried, and soliton state (red zone) is stably accessed by forward tuning the pump laser while keeping the auxiliary laser fixed. Thermal effect is mitigated by the introduction of the auxiliary laser. Note that the abscissa corresponds to the absolute detuning of the pump laser with respect to the cold cavity mode.

Before delving into the impact of the XPM term, we first consider the thermal compensation effect of the auxiliary laser. The experimentally captured soliton steps at the end of the thermal triangle are shown in Figure 3(a). It is seen that the soliton step duration is extended conspicuously from ~50 ns to ~25 μs with the introduction of the auxiliary laser, indicating effective suppression of thermal instability. Thermal compensation effect of the auxiliary laser originates from reduced variation in intracavity power during transition to soliton states [12–14]. We have calculated the intracavity comb power based on [Equation (1)], as shown in Figure 3(c). In the simulation, the frequency of the auxiliary laser is fixed and remains blue-detuned with respect to the hot cavity mode. As the pump laser is swept across the cavity mode by forward tuning, soliton steps occur at the edge of the thermal triangle. It is evident that the reduction in pump power at the soliton steps is compensated by the increase in auxiliary laser power, thus the total average intracavity power remains roughly unvaried. Since a stable intracavity power implies a stable microresonator temperature, thermal effect is effectively mitigated by the auxiliary laser, and solitons can be stably accessed by forward tuning the pump laser.

Next, we consider the XPM enabled soliton switching. For the moment, we shall assume that the thermal effect is negligible during the transition to soliton state, as a result of the thermal compensation effect of the auxiliary laser. Thus the thermal detuning $\delta_T$ can be taken as a constant during the soliton switching process. In the following simulations, $\delta_T$ is assumed to be zero for the sake of simplicity. Similar to the Kerr effect induced pump transmission[30], the XPM term in [Equation (1)] would lead to an additional pump transmission as

$$\tilde{F}_i = \frac{\mathcal{F}\left\{2\left|E_j(t,\theta)\right|^2 E_i(t,\theta)\right\}_l}{\mathcal{F}\left\{E_i(t,\theta)\right\}_l} \quad (2)$$

where $\mathcal{F}$ denotes Fourier transformation and the subscript $l$ denotes the component at the pump or the auxiliary laser frequency. Note that the imaginary part of [Equation (3)] is identically zero, as XPM does not induce any energy transfer. Thus the introduction of the auxiliary laser would result in an XPM induced frequency shift $\tilde{F}_i$. Consequently, the XPM term in [Equation (1)] is seen to act as an additional frequency shift, such that the effective detuning of the pump or auxiliary laser is

$$\Delta_i = \tilde{\delta}_i - \tilde{F}_i \quad (3)$$

where $\tilde{\delta}$ denote the absolute detuning with respect to the cold cavity resonance. It is known that the upper boundary of the soliton existence range $\Delta^{max}$ is degenerate with respect to the soliton number $N$, i.e. $\Delta^{max}(N) = \Delta^{max}(N-1) = \Delta^{max}(1)$. Thus for different, i.e. different intracavity power, the maximum absolute detuning is also different.

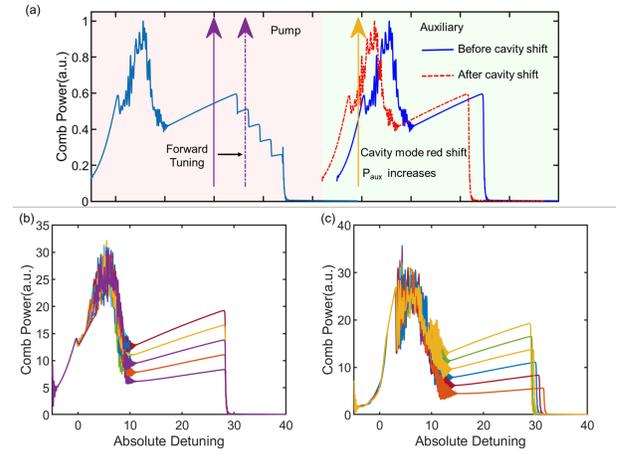

Figure 4 (a) Schematic illustration of the soliton degeneracy-lifting in a dual-pumped microresonator. Red zone: pump laser, green zone: auxiliary laser. When forward tuning the pump laser over the maximum extent of the soliton existence range, power reduction due to the loss of one soliton triggers a "chain-reaction" which would lift the soliton degeneracy. (b-c) Simulation results based on the coupled LLEs, with auxiliary laser detuning set to (b) −15 and (c) 0. The soliton remains basically degenerate when the auxiliary laser is far from resonance, but becomes nondegenerate as the auxiliary laser approaches the resonance. And soliton switching is attributed to this XPM induced degeneracy-lifting.

Now we can derive the XPM induced degeneracy-lifting from [Equation (3)], and this process is schematically illustrated in Figure 4(a). The intracavity power for $N$-soliton state is roughly proportional to the soliton number $N$. Since the absolute detuning of the auxiliary laser is fixed, the effective detuning of the auxiliary laser $\Delta_a$ would vary with the soliton number within the cavity, as the soliton induced $\tilde{F}_i$ varies with the intracavity power. Thus whenever a soliton is lost, i.e. $N \to N-1$, the XPM induced shift decreases ($\tilde{F}_a \downarrow$). As a result, the effective detuning of the auxiliary laser increases ($\Delta_a \uparrow$), and the intracavity power related to the auxiliary laser increases. The increased auxiliary laser power in turn enhances the XPM induced shift for the pump laser ($\tilde{F}_P \uparrow$), and thus decreases the effective detuning of the pump laser ($\Delta_p \downarrow$). This "chain-reaction" allows the effective detuning of the pump laser to remain within the soliton existence range. Thus the maximum absolute detuning is extended with decreasing soliton number, i.e.

$$\tilde{\delta}_p^{max}(N) < \tilde{\delta}_p^{max}(N-1) \tag{4}$$

where $\tilde{\delta}^{max}(N)$ is the upper boundary for N-soliton state. From this point of view, the cavity mode undergoes a red-shift as the soliton power decreases in the dual-pumped microresonator, corresponding to an effective negative thermo-optic effect.

Furthermore, we verify the above analysis by a full numerical simulation based on the coupled LLEs [Equation (1)], and the comb power as a function of the normalized pump laser detuning is shown in Figure 4(b-c). If the detuning of the auxiliary laser is far from the resonance (e.g. auxiliary laser detuning is –15), XPM has negligible impact on the soliton existence range, which remains degenerate with respect to soliton number. However, the soliton existence range gradually becomes non-degenerate as the auxiliary laser approaches the resonance frequency (e.g. auxiliary laser detuning is 0), as shown in Figure 4(c). Such degeneracy-lifting leads to deterministic soliton switching, in agreement with our experimental results.

We have also carried out simulations taking the thermal effects into consideration. Contrary to the XPM induced cavity mode shift, thermal effects tend to shift the cavity mode in the opposite direction. Nevertheless, it turns out that thanks to the thermal compensation effect of the auxiliary laser, XPM induced shift dominates in the dual-pumped AlN microring, manifesting an effective negative thermo-optic coefficient (Supplement 1).

In conclusion, we have demonstrated soliton switching and deterministic single soliton generation in a dual-pumped AlN microring by forward tuning the pump laser. The introduction of the auxiliary laser helps compensate the thermal effects effectively, thus allowing stably accessing soliton states without fast control of the pump laser detuning. Furthermore, we attribute the soliton switching process to the XPM induced soliton existence range degeneracy-lifting. In contrast to the thermally induced degeneracy-lifting, the soliton existence range in a dual-pumped microresonator is red-shifted as the soliton number inside the cavity decreases, which allows deterministic generation of single-soliton state by forward tuning the pump laser. The XPM induced degeneracy-lifting is confirmed by full numerical simulations based on the coupled LLEs, which are consistent with our experimental observations. The dual-pumped AlN microresonator manifests an effect negative thermo-optic coefficient, and spontaneous transition to soliton state is also recorded. This broaden the insight into the soliton dynamics in microresonators.

## ASSOCIATED CONTENT

Supporting Information Supplement 1 Available: This material is available free of charge via the Internet at http://pubs.acs.org

## AUTHOR INFORMATION

**Corresponding Author**

* E-mail: czsun@tsinghua.edu.cn

**Notes**

The authors declare no competing financial interest.

**Funding Sources**

This work was supported in part by National Key R&D Program of China (2018YFB2201700); National Natural Science Foundation of China (61975093, 61927811, 61822404, 61974080, 61904093, and 61875104); Tsinghua University Initiative Scientific Research Program (20193080036).

## ACKNOWLEDGMENT

The authors would like to thank Drs. Wenfu Zhang and Weiqiang Wang of Xi'an Institute of Optics and Precision Mechanics (XIOPM), Chinese Academy of Sciences (CAS) for their help in the experiment. And the authors would like to acknowledge helpful discussion with Prof. Hairun Guo of Shanghai University.